\documentclass{amsart}
\usepackage[latin2]{inputenc}
\usepackage{amssymb,amsthm,euscript,amsmath,graphicx}

\newcommand{\be}{\begin{equation}}
\newcommand{\ee}{\end{equation}}
\newcommand{\ben}{\begin{equation*}}
\newcommand{\een}{\end{equation*}}
\newcommand{\bg}{\begin{gather}}
\newcommand{\eg}{\end{gather}}
\newcommand{\bea}{\begin{eqnarray}}
\newcommand{\eea}{\end{eqnarray}}
\newcommand{\eean}{\end{eqnarray*}}
\newcommand{\bean}{\begin{eqnarray*}}
\newcommand\re[1]{(\ref{#1})}

\newcommand\al{\alpha}
\newcommand\ga{\gamma}
\newcommand\Ga{\Gamma}

\begin{document}
\title{Relativistic hydrodynamics - causality
and stability}

\author{P\'eter V\'an$^1$}
\address{$^1$KFKI Research Institute for Particle and Nuclear Physics
 Budapest and\\
 BCCS Bergen Computational Physics Laboratory, Bergen
}
\author{Tam\'as S. Bir\'o$^2$
} \address{$^2$KFKI Research Institute for Particle and Nuclear
Physics, Budapest}

\email{vpethm@rmki.kfki.hu}

\date{\today}

\begin{abstract}
Causality and stability in relativistic dissipative hydrodynamics
are important conceptual issues. We argue that causality is not
restricted to hyperbolic set of differential equations. E.g. heat
conduction equation can be causal considering the physical validity
of the theory. Furthermore we propose a new concept of relativistic
internal energy that clearly separates the dissipative and
non-dissipative effects. We prove that with this choice we remove
all known instabilities of the linear response approximation of
viscous and heat conducting relativistic fluids. In this paper the
Eckart choice of the velocity field is applied.
 \end{abstract}

\maketitle


\section{Introduction}

The theories of dissipative fluids are different in the relativistic
and nonrelativistic spacetime. The simplest first order, parabolic
theory of nonrelativistic fluids, the system of Navier-Stokes and
Fourier equations, is tested and justified  by countless
applications of the everyday engineering practice. The second order
generalization of the theory introduces the fluxes of the extensives
as independent variables. In this way the validity and applicability
of the hydrodynamic and heat conduction equations is extended
providing a hyperbolic system \cite{JouAta92b,MulRug98b}. Although
the well known special relativistic generalization of the first
order Navier-Stokes-Fourier system is straightforward
\cite{Eck40a3}, it has some unacceptable features. It is acausal and
unstable. Therefore it is generally accepted that only the extended,
second order theories are viable. Nowadays, heavy ion collision
experiments give a unique opportunity to check the different
suggestions and the interest in dissipative relativistic theories is
renewed \cite{HeiAta06a,KoiAta06m},
\cite{ElzAta01a,KodAta06m,BaiRom06m,BaiAta06m,Mur04a,MurRis04m,Mur06m}.

The first order theories are based on the local equilibrium
hypothesis, where the independent variables are the same as in
equilibrium, but in second (and higher) order theories the fluxes of
the local equilibrium theory appear as independent variables. An
other usual property is that in case second order relativistic
theories the entropy vector is quadratic in the fluxes, containing
terms like $q_\alpha q^\alpha u^\mu$, $\Pi^{\mu \alpha} q_\alpha$
etc., characterizing the deviation from the local equilibrium. As
one can introduce general dynamic variables beyond the fluxes (see
e.g. \cite{Ott05b}), the above classification is not general.

Differential equations of first order theories are parabolic,
therefore they are considered generally as acausal. The differential
equations of the second order theories, that are constructed
according to the Second Law, are mostly hyperbolic, therefore they
are considered generally as causal. However, more careful
considerations show, that the relation between parabolicity and
causality is not so straightforward and requires some attention both
from a mathematical and from a physical point of view.

The homogeneous equilibrium in first order theories is generally
considered as unstable. The homogeneous equilibrium of second order
theories is generally considered as stable. The stability
considerations are referring to the linear stability calculations of
Hiscock and Lindblom \cite{HisLin85a,HisLin87a}.

However, like these second order theories are extending the validity
of the first order ones, their physical content is included in the
corresponding more general second order theory. The more involved
second order theories do not cure necessarily the instabilities of
the first order theories. It is shown by Geroch and Lindblom that
physical fluid states in these theories relax to the solutions of
the underlying first order theory \cite{Ger95a,Lind96a}.

The missing of a simple and stable relativistic generalization of
the Navier-Stokes-Fourier theory resulted in several attempts to
improve the properties of first order theories.
Garc\'\i{}a-Col\'\i{}n and Sandoval-Villalbazo suggested a
separation of an internal energy balance from the balance of the
energy-momentum, similarly to nonrelativistic theories
\cite{GarSan06a}. However, with an additional independent energy
balance the energy-momentum tensor would not embrace the whole
energy content of the matter \cite{BorChr05m}. Other authors suggest
a suitable definition of the four-velocity field \cite{TsuAta06m}.
None of the previous suggestions investigate the stability of the
corresponding equations.

In the following section we argue that the speed of the propagation
of signals can be finite in parabolic theories, too, if their
physical validity is considered. Therefore first order relativistic
hydrodynamic theories cannot be excluded by referring to causality.
Moreover, in the light of the above mentioned observation of Geroch
and Lindblom it is even more important to find a viable relativistic
generalization of the Navier-Stokes-Fourier equations. A necessary
condition for a causality in a weaker sense \cite{Cim04a} is the
stability of the homogeneous solutions of the corresponding
differential equations. Based on this observation in the followings
we outline a new approach to relativistic fluids. We suggest a
separation of the dissipative and non-dissipative parts of the
energy momentum distinguishing between the total energy density and
the internal energy density of the matter. The later is the absolute
value of the projected energy flux four-vector, this way it
incorporates the momentum density as well. Since in the
corresponding thermodynamic frame the entropy density depends also
on the energy flux besides the energy density, but does not on the
pressure, our suggestion can be classified between the first order,
local equilibrium one, and of the extended, second order theories.
In the final section we demonstrate the linear stability of the
homogeneous equilibrium of viscous, heat conducting fluids with the
Eckart choice of the velocity field.

\section{Remarks on causality}

The common argument against the use of parabolic differential
equations in physics is that some of their typical solutions show
signal propagation with infinite speed. More sophisticated arguments
require a well posedness of the related mathematical problems that
can be guaranteed by hyperbolicity. The characteristic surface of
the simplest relativistic heat conduction equation with constant
coefficient is a spacelike hypersurface according to a comoving
observer. Moreover, the characteristic surfaces are invariant to the
transformation of the equation (in particular to a Lorentz boost of
the reference frame): The parabolicity or the hyperbolicity of the
equation does not change by changing the observer. Therefore, as it
was argued by Kost\"adt and Liu \cite{KosLiu00a}, by simple
mathematics initial value problems of parabolic differential
equations can be well posed, provided that initial data are given on
the characteristic surface of the equations. From a physical point
of view this is a natural requirement.

On the other hand, the characteristic hypersurfaces are those that
determine the speed of propagation of simple solutions (the domain
of influence) of a hyperbolic differential equation, too. Therefore,
speed of the signal propagation for a hyperbolic differential
equation is in general not infinite, but can be higher than the
speed of light. The actual speed depends on the parameter values in
the equation. This statement appears as a trivial fact in case of
wave propagation equation. In one space dimension, considering a
comoving observer, with respect to a constant velocity field $u_\al$
we get the following form
 \be
\partial_{tt} \theta - c_w^2 \partial_{xx}\theta = 0.
\label{we}\ee

Here $\theta$ is the corresponding scalar physical quantity, $c_w$
is the wave propagation speed. The solution of the characteristic
differential equation of \re{we} gives the equation $\theta(x,t) = x
\pm c_w t = const.$ for the two characteristic lines of the
equation. Applying a Lorentz transformation $\tilde{x} = \ga (x-vt)$
and $\tilde{t} = \gamma(t-vx/c^2)$ we can get the transformed form
of the characteristic lines as:
$$
\theta(\tilde{x},\tilde{t}) = (1\pm \frac{c_w v}{c^2})x + (v\pm
c_w){t}.
$$

Therefore the transformed characteristic speed is
$$
  \tilde{c}_w = \frac{v\pm c_w}{1\pm \frac{c_w v}{c^2}}.
$$

We can get the same result with the Lorentz transformation of the
equation, too. The above expression shows that the propagation speed
of waves can be faster than the speed of the light only if $c_w
> c$, as we have expected.

In case of a set of nonlinear differential equations the calculation
of the characteristic wave speeds can be more involved and even the
proof of the hyperbolicity of the corresponding set of equations is
not trivial. In general the value of the speed will depend on
parameters in the set of equations and the relativistic, covariant
form combined with hyperbolicity do not warrant a propagation speed
smaller than the speed of light.

On the other hand the theoretically infinite speeds in parabolic
equations are usually not observable, because their effect is out of
the physical validity range of the theory. The atomistic structure
of the matter restricts the validity of continuum descriptions as it
was pointed out by Weynmann \cite{Wey67a,KosLiu00a}. With the help
of the mean free path and the collision time one can give simple
estimates on the propagation speed of measurable signals.

We demonstrate this property on the example of the Fourier heat
conduction equation
 \be
 \partial_t \theta - \lambda \partial_{xx} \theta = 0.
\ee

The hydrodynamic range of validity requires that $\theta$ must not
vary too rapidly over a mean free path $\xi$
 \be
  \left|
\frac{1}{\theta} \frac{\partial \theta}{\partial x} \right| <<
\frac{1}{\xi}.
 \label{cval}\ee

Assuming a sharp initial condition the solution of the heat
conduction equation can be written as
$$ \theta(x,t) =
\frac{A}{\sqrt{2\pi t}} e^{-\frac{x^2}{4 \lambda t}}
$$

This is a typical acausal solution of the Fourier equation. However,
substituting the above solution into the condition \re{cval} we get
a limit of the propagation speed of the continuum signals as
 \be
 \frac{x}{t} << v_{lim} \propto \frac{\lambda}{\xi}.
 \ee

Therefore, instead of the infinite tail of the solution in the
reality we have an extending range, cf.  \re{Fig1}.

\begin{figure}\centering
\includegraphics{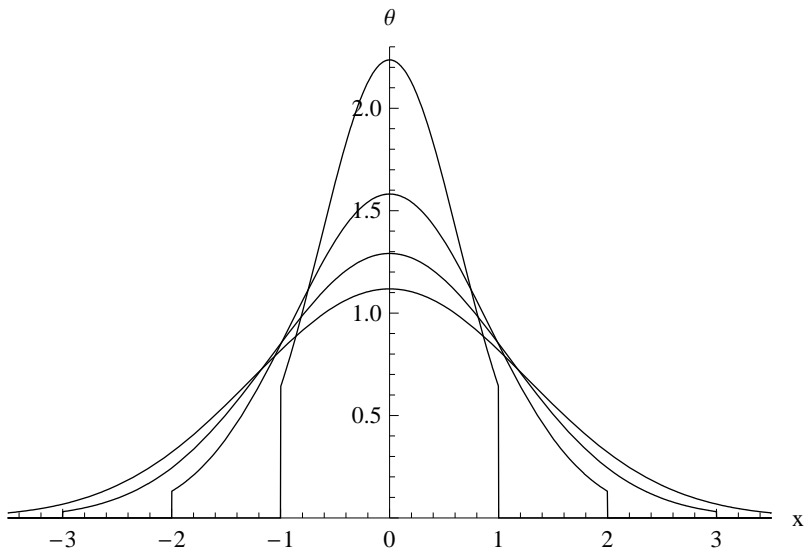}
\caption{Mean free path limited signal propagation according to the
Fourier heat conduction equation. $\xi= .2$, $\lambda=1$ and
$t=0.2,0.4,0.6,0.8$} \label{Fig1}
\end{figure}

In case of heat conduction in water at room temperature we can
easily give an estimate as $v_{max} \sim \frac{\Lambda}{c_v \rho\xi}
\simeq 14 m/s$, where $\Lambda$ is the Fourier heat conduction
coefficient, $\rho$ is the density and $c_v$ is the specific heat of
water. As heat conduction is disputed phenomena in quark-gluon
plasma we cannot give a reliable estimation here.

Independently of the previous estimation Fichera suggested that the
speed of the signal propagation is restricted by observability of
the given physical quantity. Observability can be related to the
sensitivity of the measurement but also to fluctuations and the
particular structure of the matter. That can give an other bound to
the speed of the signal propagation \cite{Fic92a,Day97a,Cim04a}. In
our case we may assume that we cannot observe $|\theta|$ below a
given value $|\theta| < \theta_{max}$. Then the propagation speed
becomes finite, nevertheless it is not constant as one can inspect
in figure \re{Fig2}.

\begin{figure}\centering
\includegraphics{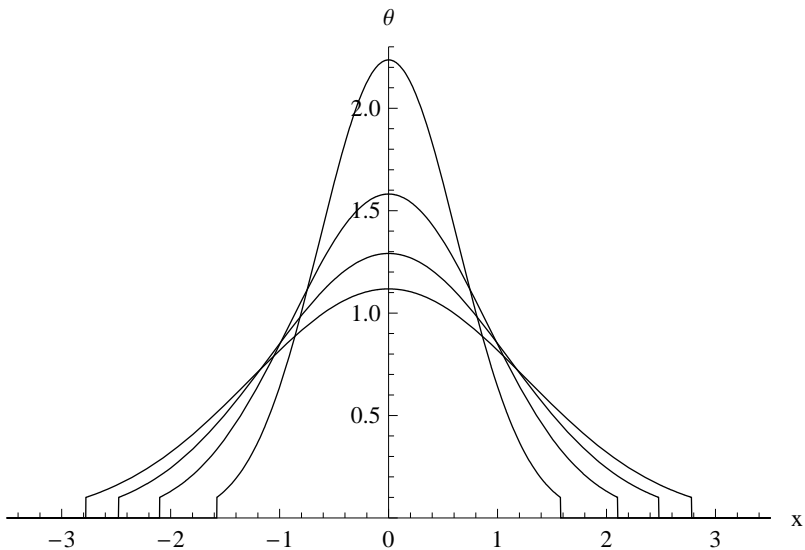}
\caption{Signal propagation according to the observability threshold
of the Fourier heat conduction equation. $\theta_{max}= 0.1$,
$\lambda=1$ and $t=0.2,0.4,0.6,0.8$}. \label{Fig2}
\end{figure}

Summarizing our arguments have seen that hyperbolicity of the
equations can lead to well posed problems and gives finite
propagation speed but does not warrant that the propagation speed is
less than the speed of light. On the other hand one can formulate
well posed Cauchy problems related to parabolic equations, too.
Moreover, the physical validity of a continuum theory can warrant
slow propagation speeds in several different ways. Therefore, we may
conclude that parabolic and mixed systems of continuum differential
equations (as Fourier heat conduction or any first order continuum
hydrodynamics) could be useful models in relativistic theories.
Those theories cannot be excluded by causality arguments.

All the previous estimates are connected to some definite properties
of the solutions of the simple heat conduction equation. If the
exponential damping of the solutions cannot be guaranteed, then
causality issues can become important. Therefore the stability of
the homogeneous equilibrium is not only an evident physical
requirement but also a necessary condition for the causality of any
first order dissipative relativistic hydrodynamics.

\section{Balances of particle number, energy and momentum of
relativistic fluids}

For the metric (Lorentz form) we use the $g^{\mu\nu}=
diag(-1,1,1,1)$ convention and we use a unit speed of light $c=1$,
therefore for a four-velocity $u^\al$ we have $u_\al u^\al = -1$.
$\Delta^\al_{\;\;\beta} = g^\al_{\;\;\beta} + u^\al u_\beta$ denotes
the $u$-orthogonal projection. This metric convention will be
convenient in the stability calculations.

In the following we fix the velocity field to the particle number
flow, according to Eckart. Therefore the particle number flow is
timelike by definition and can be expressed by the local rest frame
quantities as
 \be
 N^\alpha = n u^\alpha.
 \label{N}\ee

Here $n= -u_\alpha N^\alpha$ is the {\em particle density} in
comoving frame.

The particle number conservation is described by
 \be
 \partial_\alpha N^\alpha = \dot{n} + n\partial_\alpha u^\alpha = 0,
 \label{nbal}\ee

\noindent where $\dot{n} = \frac{d n}{d\tau} = u^\alpha
\partial_\alpha n$ denotes the derivative of $n$ with respect to the proper time
$\tau$.

The energy-momentum density tensor is given with the help of the
rest-frame quantities as
 \be
  T^{\alpha\beta} = e u^\alpha u^\beta + u^\alpha q^\beta +
    u^\beta {q}^\alpha + P^{\alpha\beta},
 \label{T}\ee

\noindent where $e = u_\alpha u_\beta T^{\alpha\beta}$ is the {\em
density of the energy}, $q^\beta =
-u_\alpha\Delta^\beta_{\;\;\gamma} T^{\alpha\gamma}$ is the  {\em
energy flux} or {\em heat flux}, ${q}^\alpha =
-u_\beta\Delta^\alpha_{\;\;\gamma}T^{\gamma\beta}$ is the {\em
momentum density} and $P^{\alpha\beta}= \Delta^\alpha_\gamma
\Delta^\beta_\mu T^{\gamma\mu}$ is the {\em pressure (stress)
tensor}. The momentum density, the energy flux and the pressure are
spacelike in the comoving frame, therefore $u_\alpha q^\alpha = 0$
and $u_\al P^{\alpha\beta} = u_\beta P^{\alpha\beta} = 0^\beta$. Let
us emphasize that the \re{T} form of the energy-momentum tensor is
completely general, it is just expressed by the local rest frame
quantities. The energy-momentum tensor is symmetric, because we
assume that the internal spin of the material is zero. In this case
the heat flux and the momentum density are equal. However, the
difference in their physical meaning is a key element of our train
of thoughts. Heat is related to dissipation of energy but momentum
density is not, therefore this difference should appear in the
corresponding thermodynamic framework.

Now the conservation of energy-momentum  $\partial_\beta
T^{\alpha\beta} = 0$ is expanded to \be
 \partial_\beta T^{\alpha\beta} =
    \dot{e} u^\al + e u^\al \partial_\beta u^\beta +e\dot{u}^\al +
    u^\al\partial_\beta q^\beta +q^\beta \partial_\beta u^\al +
    \dot{{q}}^\al + {q}^\al \partial_\beta u^\beta +
    \partial_\beta P^{\al\beta}.
\label{E-Ibal}\ee

Its timelike part in the local rest frame gives the balance of the
energy $e$
 \be
    -u_\alpha\partial_\beta T^{\alpha\beta} =
        \dot{e} + e \partial_\alpha u^\alpha +
        \partial_\alpha q^\alpha + {q}^\alpha \dot{u}_\alpha +
        P^{\al\beta} \partial_\beta u_\alpha = 0.
 \label{ebal}\ee

The spacelike part in the local rest frame describes the balance of
the momentum
 \be
    \Delta^\alpha_{\;\;\gamma}\partial_\beta T^{\gamma\beta} =
        e\dot{u}^\alpha  +
        {q}^\alpha \partial_\beta u^\beta +
        q^\beta \partial_\beta u^\alpha +
        \Delta^\alpha_{\;\;\gamma} \dot{{q}}^\ga +
        \Delta^\al_\ga\partial_\beta P^{\beta\ga}
    = 0.
 \label{ibal}\ee

\section{Thermodynamics}

The entropy density and flux can also be combined into a
four-vector, using local rest frame quantities:
 \be
  S^\alpha = s u^\alpha + J^\alpha,
 \label{S}\ee
\noindent where $s= -u_\alpha S^\alpha$ is the  {\em entropy
density} and $J^\alpha = S^\alpha - u^\alpha s =
\Delta^\alpha_{\;\;\beta} S^\beta$ is the {\em entropy flux}. The
entropy flux is $u$-spacelike, therefore $u_\alpha J^\alpha = 0$.
Now the Second Law of thermodynamics is translated to the following
inequality
 \be
 \partial_\alpha S^\alpha = \dot{s} + s\partial_\alpha u^\alpha +
    \partial_\alpha J^\alpha \geq 0
 \label{sbal}\ee

Relativistic thermodynamic theories assume that the entropy is a
function of the local rest frame quantities, because the
thermodynamic relations reflect general properties of local material
interactions. The most important assumption is that the entropy is a
function of the local rest frame energy density, the time-timelike
component of the energy momentum tensor according to the velocity
field of the material \cite{Eck40a3,Tol34b}. Definitely the
thermodynamics cannot be related to an external observer, therefore
the dependence on the relative kinetic energy is excluded. This
interpretation of $e$ in \re{T} is supported by the form of the
energy balance \re{ebal}, where the last term is analogous to the
corresponding internal energy source (dissipated power) of the
nonrelativistic theories.

In nonrelativistic fluids the internal energy is the difference of
the conserved total energy and the kinetic energy of the material.
However, also in nonrelativistic theories the constitutive relations
must be objective in the sense that they cannot depend on an
external observer, the thermodynamic framework should produce frame
independent material equations. (This apparent contradiction of
classical physics is eliminated by different sophisticated methods
and lead to such important concepts as the configurational forces
or/and virtual power \cite{Mau99b,Gur00b,Sil97b}). However, without
distinguishing the energy related to the flow of the material from
the total energy one mixes the dissipative and nondissipative
effects. The wrong separation leads to generic instabilities of the
corresponding theory.

Our candidate of the relativistic internal energy is related to the
energy vector defined by $E^\al = -u_\beta T^{\al\beta} = e u^\al +
q^\al$. The energy vector embraces both the total rest frame energy
density and the rest frame momentum. Therefore its absolute value
$\|E\|= \sqrt{-E_\al E^\al} = \sqrt{|e^2 - q_\al q^\al|}$ seems to
be a reasonable choice of the scalar internal energy\footnote{Note
that $q^\al$ is spacelike, therefore the quantity under the sign is
non-negative.}. Its series expansion, when the energy density is
larger than the momentum density, shows strong analogies to the
corresponding nonrelativistic definition
 $$
  \|E\| =\sqrt{|e^2 - q_\al q^\al|} \approx e -
    \frac{{\bf q}^2}{2 e} + ...
 $$

Thermodynamic calculations based on the Liu procedure support this
assumption \cite{Van07m2}. Let us emphasize, that our candidate of
internal energy is not related to any external reference frame, only
to the velocity field of the material. In a Landau-Lifschitz frame
the energy vector is timelike.

Assuming that the entropy density is the function of the internal
energy and the particle number density
$s(e,q^\al,n)=\hat{s}(\sqrt{|e^2 - q_\al q^\al|},n)$ leads to a
modified form of the thermodynamic Gibbs relation and the potential
relation for the densities as follows
 \be
  de - \frac{q^\al}{e} dq_\al = T  ds  + \mu dn,
  \qquad \text{and} \qquad
  e - \frac{q^2}{e} = T s - p + \mu n.
\label{eqt}\ee

Here $T$ is the temperature, $p$ is the pressure and $\mu$ is the
chemical potential. Equivalently the Gibbs relation gives the
derivatives of the entropy density as follows
 $$
 \left. \frac{\partial s}{\partial e}\right|_{(q^\al,n)}
    = \frac{1}{T}, \qquad
 \left. \frac{\partial s}{\partial q^\al}\right|_{(e,n)}
    = -\frac{q_\al}{eT}, \qquad
  \left. \frac{\partial s}{\partial n}\right|_{(q^\al,e)}
 = -\frac{\mu}{T}.
$$

For the entropy flux we assume the classical form
 \be
 J^\alpha = \frac{q^\al}{T }.
 \label{sf}\ee

Now we substitute the energy balance \re{ebal} and the particle
number balance \re{nbal} into the entropy balance \re{sbal} and we
arrive at the following entropy production formula:
 \bea
  \partial_\alpha S^\alpha &=&
    \dot{s}(e,q^\al,n) +
    s \partial_\alpha u^\alpha +
    \partial_\alpha J^\al \nonumber \\
  &=& \frac{\partial s}{\partial e}\dot{e} +
    \frac{\partial s}{\partial q^\al} \dot{q}^\al +
    \frac{\partial s}{\partial n}\dot{n} +
    s \partial_\alpha u^\alpha +
    \partial_\alpha \frac{q^\al}{T} \nonumber \\
  &=& -\frac{1}{T }(e\partial_\alpha u^\alpha +
        \partial_\alpha q^\alpha +
        q^\al \dot{u}_\al +
        P^{\al\beta}\partial_\beta u_\al) -
    \frac{q^\al}{T e}\dot{q}_\al +
    s\partial_\alpha u^\alpha  \nonumber\\
  &+& \frac{\mu}{T} n \partial_\al u^\al +
    q^\al \partial_\alpha\frac{1}{T } +
    \frac{1}{T} \partial_\al q^\al  \nonumber\\
  &=& -\frac{1}{T }\left(P^{\alpha\beta} -
    (-e+sT+\mu n)\Delta^{\alpha\beta}\right)
        \partial_\alpha u_\beta  +
    {q}^\alpha\left(
        \partial_\alpha\frac{1}{T}
        -\frac{\dot{u}_\al}{T} - \frac{\dot{q}_\al}{e T}
       \right)\geq 0.
 \label{cld}\eea

According to this quadratic expression and the potential relation in
\re{eqt} the {\em viscous pressure} is given by
$$
\Pi^{\al\beta} = P^{\al\beta}- \left(p-\frac{q^2}{e}
\right)\Delta^{\al\beta}
$$

We may introduce the conventional decomposition of the pressure
$$
  \Pi^{\al\beta}= (p+\Pi) \Delta^{\alpha \beta} +
  \langle\Pi^{\alpha\beta}\rangle.
$$
Here $\Pi =
\frac{1}{3}\Pi^\al_\al-p=\frac{1}{3}P^\al_\al-\frac{q^2}{e}$ and
$\langle\Pi^{\al\beta}\rangle= \Delta^\al_\mu\Delta^\beta_\nu
(\frac{1}{2}(\Pi^{\mu\nu}+\Pi^{\nu\mu}) - \frac{1}{3}
\Delta^{\mu\nu}\Delta_{\ga\delta} \Pi^{\ga\delta}$ is symmetric an
traceless.

Therefore the \re{cld} entropy production can be written as
 \be
 -\frac{1}{T}\langle\Pi^{\alpha \beta}\rangle\partial_\alpha u_\beta
 -\frac{1}{T}\left(\Pi - \frac{q^2}{e}\right)\partial_\alpha u^\al
 +q^\alpha\left(\partial_\alpha\frac{1}{T}-
    \frac{\dot{u}_\alpha}{T}
    -\frac{\dot{q}_\alpha}{eT}\right) > 0
\label{entrpr}\ee

In isotropic continua the above entropy production results in the
following constitutive functions assuming a linear relationship
between thermodynamic fluxes and forces
 \bea
  q^\al &=&
    -\tilde\lambda \frac{1}{T^2} \Delta^{\alpha\gamma}
    \left(\partial_\gamma T +
        T \dot{u}_\ga +
        \frac{\dot{q}_\ga}{e} \right),\label{fo}\\
  \langle\Pi^{\alpha\beta}\rangle &=&
    -2\eta \langle\partial^\al u^\beta\rangle, \label{newts}\\
  \Pi -\frac{q^2}{e} & = & -\eta_v \partial_\alpha u^\al.\label{newtb}
\eea

Let us recognize the additional term to the bulk viscous pressure.
For the whole viscous stress we get
 \be
 \Pi^{\alpha\beta} = -2\eta \langle\partial^\al u^\beta\rangle
 -\eta_v \Delta^{\alpha \beta}\partial_\ga u^\ga.
 \label{newt}\ee

\re{fo} and \re{newt} are the relativistic generalizations of the
Fourier law of heat conduction and the Newtonian viscous pressure
function. The shear and bulk viscosity coefficients, $\eta$ and
$\eta_v$ and the heat conduction coefficient $\lambda=
\tilde{\lambda}T^{-2}$ are non negative, according to the inequality
of the entropy production \re{fo}. We may introduce a relaxation
time $\tau = \lambda/e$ in \re{fo}, as usual in second order
theories.

The equations \re{nbal} and \re{ebal} are the evolution equations of
a relativistic heat conducting ideal fluid, together with the
constitutive function \re{newt} and the relaxation type equation
\re{fo}. As special cases we can get the relativistic Navier-Stokes
equation substituting \re{newt} into \re{ibal} and assuming $q^\al =
0$, or the equations of relativistic heat conduction solving
together \re{fo} and \re{ebal} assuming that $\Pi^{\al\beta}=0$ and
$u^\al = const.$.

\section{Linear stability}

In this section we investigate the linear stability of the
homogeneous equilibrium of the equations \re{nbal}, \re{ebal} and
\re{ibal} together with the constitutive relations \re{newt} and
\re{fo}. Similar calculations are given by Hiscock and Lindblom both
for Eckart fluids \cite{HisLin85a} and Israel-Stewart fluids
\cite{HisLin87a}.

\subsection{Equilibrium}

The equilibrium of the above set of equations is defined by
vanishing proper time derivatives and by zero entropy production
with vanishing thermodynamic fluxes
 \be
 \Pi^{\alpha\beta} = 0  \quad \textrm{and} \quad q^a = 0.
\label{et1}\ee

Therefore according to the balances and the constitutive functions
the equilibrium of the fluid is determined by
 \begin{gather}
 n = \text{const.} \quad e = \text{const.} \quad \Rightarrow
  \quad T= \text{const.}, \quad \mu = \text{const.}, \quad
    p= \text{const.}, \label{eq1}\\
  \partial_\alpha u_\alpha = 0, \qquad
    \partial_\alpha u_\beta + \partial_\beta u_\alpha =0. \label{eq2}
 \end{gather}

In addition to the above conditions we require a homogeneous
equilibrium velocity field
 \be
  u_\alpha = \text{const.}
 \label{e3}\ee

\subsection{Linearization}

We denote the equilibrium fields by zero lower index and the
perturbed fields by $\delta$ as $Q = Q_0 +\delta Q$. Here $Q$ stands
for $n$, $e$, $u^\al$, $q^\al$, and $\Pi^{\al\beta}$. The linearized
equations \re{nbal}, \re{ebal}, \re{ibal}, \re{fo}, \re{newt} around
the equilibrium given by \re{et1}-\re{eq1}-\re{e3} become
 \bea
 0 &=& \dot{\delta n} +
        n \partial_\alpha \delta u^\alpha, \label{lnbal}\\
 0 &=& \dot{\delta e} +
        (e+p) \partial_\alpha \delta u^\alpha +
        \partial_\alpha \delta q^\alpha, \label{lteb}\\
 0 &=& (e+p)\dot{\delta u^\al}  +
        \Delta^{\alpha\beta} \partial_\beta \delta p +
        \dot{\delta q^\al} +
        \Delta^\al_\ga \partial_\beta \delta\Pi^{\ga\beta},
        \label{ltib}\\
 0 &=& \delta q^\alpha +
        \lambda \Delta^{\alpha\gamma}\left(\partial_\gamma \delta{T} +
        T \dot{\delta u}_\ga +
        \frac{T}{e} \dot{\delta q}_\ga\right), \label{lfum}\\
 0 &=&  \delta\Pi^{\alpha\beta} +
        \tilde{\eta}_v \partial_\gamma \delta u^\gamma \Delta^{\alpha\beta} +
        \eta \Delta^{\alpha\gamma}\Delta^{\beta\mu}(
            \partial_\gamma\delta u_\mu +\partial_\mu\delta u_\ga).
        \label{lnewtm}
\eea

Here $\tilde{\eta}_v=(\eta_v-\frac{2}{3}\eta)$. The perturbation
variables satisfy the following properties inherited from the
linearization of the original ones
 $$
 0 = u^\al \delta q_\al =
 u^\al \delta u_\al =
 u^\al \delta \Pi_{\al\beta} =
 \delta \Pi_{\al\beta} -  \delta \Pi_{\beta\al}
$$

In order to identify possible instabilities we select out
exponential plane-wave solutions of the perturbation equations:
$\delta Q =  Q_0 e^{\Gamma t + i k x}$, where $Q_0$ is constant and
$t$ and $x$ are two orthogonal coordinates in Minkowski spacetime.
As our equilibrium background state is a fluid at rest we put
$u^\al\partial_\al =\partial_t$.

With these assumptions the set of perturbation equations follow as
 \bea
 0 &=& \Ga{\delta n} + i k n \delta u^x,
    \label{lin1}\nonumber\\
 0 &=& \Ga{\delta e} +(e+p) i k \delta u^x +i k \delta q^x,
    \label{lin2}\nonumber\\
 0 &=& \Ga(e+p)\delta u^x + i k (\partial_e p \delta e +
    \partial_n p \delta n)+ \Gamma \delta q^x + i k \delta\Pi^{xx},
    \label{lin3x}\nonumber\\
 0 &=& \Ga(e+p)\delta u^y + \Gamma\delta q^y + i k \delta\Pi^{xy},
    \label{lin3y}\nonumber\\
 0 &=& \Ga(e+p)\delta u^z + \Gamma\delta q^z + i k \delta\Pi^{xz},
    \label{lin3z}\nonumber\\
 0 &=& \delta q^x + i k \lambda(\partial_e T \delta e +
    \partial_n T \delta n)+\lambda T \Gamma \delta u^x +
    \lambda \frac{T}{e} \Gamma \delta q^x,
    \label{lin4x}\nonumber\\
 0 &=& \delta q^y +\lambda T \Gamma \delta u^y +
    \lambda \frac{T}{e} \Gamma \delta q^y,
    \label{lin4y}\nonumber\\
 0 &=& \delta q^z +\lambda T \Gamma \delta u^z +
    \lambda \frac{T}{e} \Gamma \delta q^z,
    \label{lin4z}\nonumber\\
 0 &=&  \delta\Pi^{xx} + ik \tilde{\eta}\delta u^x,
    \label{lin6xx}\nonumber\\
 0 &=&  \delta\Pi^{xy} + ik \eta\delta u^y,
    \label{lin6xy}\nonumber\\
 0 &=&  \delta\Pi^{xz} + ik \eta\delta u^z,
    \label{lin6xz}\nonumber\\
 0 &=&  \delta\Pi^{yy} + ik \tilde{\eta}_v\delta u^y,
    \label{lin6yy}\nonumber\\
 0 &=&  \delta\Pi^{zz} + ik \tilde{\eta}_v\delta u^z,
    \label{lin6zz}\nonumber\\
 0 &=&  \delta\Pi^{zy}.
    \label{lin6zy}
\eea

Here we have introduced a shortened notation for $\tilde{\eta} =
\eta_v +\frac{4}{3}\eta$. We can put the equations above into the
following matrix form
 \be
  M^A_{\;\;B} \delta Q^B = 0.
\label{meq}\ee

Here $\delta Q^B$ represents the list of fields which describe the
perturbation of the fluid:
\begin{gather*}
 \delta Q =
    (\delta n, \delta e, \delta u^x, \delta q^x, \delta \Pi^{xx},\;\;
    \delta u^y, \delta q^y, \delta \Pi^{xy},\delta \Pi^{yy} \\
    \delta u^z, \delta q^z, \delta \Pi^{xz},\delta \Pi^{zz},\;\;
    \delta \Pi^{yz}).
\end{gather*}

Then the 14x14 matrix {\bf M} can be written in the block diagonal
form
 \be
  {\bf M} = \begin{pmatrix}
  {\bf N} & 0       & 0       & 0       \\
  0       & {\bf R} & 0       & 0       \\
  0       & 0       & {\bf R} & 0       \\
  0       & 0       & 0       & 1 \end{pmatrix},
 \label{M}\ee

\noindent where the submatrices {\bf R} and {\bf N} are defined as
follows
 \be
  {\bf R} = \begin{pmatrix}
  (e+p)\Ga          & \Ga                       & ik    & 0\\
  \lambda \Ga T     & 1+\lambda \Ga\frac{T}{e}  & 0     & 0\\
  ik\eta            & 0                         & 1     & 0\\
  ik\tilde{\eta}_v  & 0                         & 0     & 1\\
  \end{pmatrix},
 \label{R}\ee

\be
  {\bf N} = \begin{pmatrix}
  \Ga                   & 0                     & ikn      & 0    & 0 \\
  0                     & \Ga                   & ik(e+p)  & i k  & 0 \\
  ik\partial_n p        & ik\partial_e p        & \Ga(e+p) & \Ga  & ik \\
  ik\lambda\partial_n T & ik\lambda\partial_e T
        & \lambda \Ga T     & 1+\lambda \Ga\frac{T}{e} & 0 \\
  0                     & 0
        & ik\tilde{\eta}    & 0                        & 1
  \end{pmatrix}.
 \label{Nm}\ee

Exponentially growing plane-wave solutions of \re{meq} emerge
whenever $\Gamma $ and $k$ satisfy the dispersion relation
 \be
 \det {\bf M} = (\det {\bf N})(\det {\bf R})^2=0
\ee with a positive real $\Gamma$. The roots of this equation are
the roots obtained by setting the determinants of either {\bf N} or
{\bf R}  to zero.

The determinant of {\bf R} gives the condition
 $$
 \lambda T \frac{p}{e} \Ga^2 +
 \left(e+p+ k^2 \tilde{\eta}\lambda\frac{T}{e}\right)\Ga +
 \tilde{\eta} k^2 = 0.$$

The real parts of the roots of this polinomial are negative because
the coefficients of both the linear and the quadratic term are
positive.

The determinant of {\bf N} gives the following dispersion relation
 \begin{gather*}
 \lambda p \frac{T}{e} \Ga^4  +
 \left(e + p + k^2\tilde{\eta}\lambda\frac{T}{e}\right) \Ga^3 +
 k^2 \left(\tilde{\eta} +
        \lambda \frac{T}{e}(n \partial_n p + p \partial_ep) -
        \lambda n \partial_nT\right)\Ga^2 +\\
 k^2 \left((e +p)\partial_e p + n \partial_n p  +
    k^2 \tilde{\eta}\lambda \partial_e T  \right) \Ga  +
 k^4\lambda n \left(\partial_e T\partial_n p  -
        \partial_e p \partial_n T \right)= 0.
 \end{gather*}

According to the Routh-Hurwitz criteria \cite{KorKor00b}, the real
parts of the roots of a fourth order polynomial $a_0 x^4+a_1 x^3 +
a_2 x^2 + a_3 x + a_4=0$ are negative whenever \bea
 a_0 &>& 0, \nonumber\\
 a_1 &>& 0, \nonumber\\
 a_1 a_2 - a_0 a_3&>& 0, \nonumber \\
 (a_1 a_2 - a_0 a_3)a_2 - a_4 a_1^2 &>& 0.
\label{RH}\eea

We can see, that the first two conditions of \re{RH} are fulfilled
according to the Second Law, the nonnegativity of the entropy
production.

Let us recall the conditions of thermodynamic stability
 \bea
    \partial_e T > 0, \label{s1} \\
    \partial_n \frac{\mu}{T} > 0, \label{s2} \\
    \partial_e T \partial_n \frac{\mu}{T} -
        \partial_n T \partial_e \frac{\mu}{T} \geq 0, \label{s3}
  \eea
\noindent and the following useful identities
 \bean
    T \partial_e p &=& (e+p)\partial_e T + n T^2\partial_e \frac{\mu}{T}, \\
    T \partial_n p &=& (e+p) \partial_n T + n T^2 \partial_n \frac{\mu}{T}.
\eean

Now, the third condition can be written in a simplified form as
 \begin{gather}
  a_1 a_2 - a_0 a_3 =
  k^2 \lambda(n T)^2 \partial_n \frac{\mu}{T} +
    k^4 \lambda  \tilde{\eta}^2 \frac{T}{e}  +
    k^2 \tilde{\eta} (e+p) + \nonumber\\
  k^4 \tilde{\eta}\lambda^2 \frac{T}{e^2} \left(
    \partial_e T p^2 +
    2 n  \partial_n T p +
    T^2 n^2 \partial_n\frac{\mu}{T} \right)\geq 0.
\label{cond3}\end{gather}

The first three terms in the expression are positive. In the
parenthesis of the last term we can recognize a second order
polynomial of $p$. The discriminant of that polynomial is negative
 \begin{gather}
D_1= (2 n \partial_n T)^2 - 4 \partial_e T n^2 T^2 \partial_n
 \frac{\mu}{T} =
 - 4 n^2\left(\partial_e T \partial_n \frac{\mu}{T} -
        \partial_n T \partial_e \frac{\mu}{T}\right) < 0,
\label{d1}\end{gather}

\noindent because of the last condition of thermodynamic stability.
Therefore the expression in the parenthesis is positive for all p.

Hence the fourth condition of \re{RH} expands to the following form
\begin{gather*}
  (a_1 a_2 - a_0 a_3)a_2 - a_4 a_1^2=\\
  k^4 (e+p)\frac{\eta}{T}\left( \partial_e T (e+p)^2  +
    2 n \partial_n T  (e+p)+
    T^2 n^2 \partial_n\frac{\mu}{T} \right) +\\
  \lambda k^4 \frac{n^2}{T^2}\left(
    \partial_nT(e+p)+
    \partial_n\frac{\mu}{T}nT^2\right)^2 + \\
  \lambda k^6\frac{\eta^2}{e}\left(
    e(e+p)\partial_eT + \partial_e T (e+p)^2  +
    2 n \partial_n T  (e+p)+
    T^2 n^2 \partial_n\frac{\mu}{T}\right)+\\
  \lambda^2 k^6\frac{\eta}{e^2}(p(e+p) \partial_eT+
    n^2T^2 \partial_n \frac{\mu}{T})\left[
    (p(e+p)\partial_eT+ (e+2p)\partial_nT)^2 +
    \right.\\\left.
    n^2e(e+2p)(\partial_nT)^2 +
    n^2T^2\partial_n\frac{\mu}{T} \left(
        2p^2\partial_eT +
        n^2T^2\partial_n\frac{\mu}{T}+2n(e+2p)\partial_nT\right)\right]\\
  \lambda^2 k^8\eta^3\frac{T}{e}\partial_eT +
  \lambda^3 k^8 \eta^2 \partial{T}{e^2}(n\partial_nT+
  p\partial_eT)^2 > 0.
\end{gather*}

In the first and third term we recognize the same polynomial
expression of $(e+p)$ as in \re{cond3} for $p$. Therefore all terms
are clearly positive, only the term in the rectangular parenthesis
requires separate investigation. We may recognize that it is a
second order polynomial of $e$, with the discriminant
$$
 D = -4 n^2(n\partial_nT+p\partial_eT)^2\left(
    2p^2(\partial_eT\partial_n\frac{\mu}{T} T^2 -
    (\partial_nT)^2) -
    (p\partial_nT + nT^2 \partial_n\frac{\mu}{T})^2\right)<0.
$$

The coefficient of the $e^2$ term is $\partial_e T p^2 + 2 n
\partial_n T p + T^2 n^2 \partial_n\frac{\mu}{T}>0$ is a positive
quantity according to \re{d1}. Therefore the term in the rectangular
parenthesis is positive, too.

We conclude that the homogeneous equilibrium of the relativistic
heat conducting viscous relativistic fluids is stable in contrast to
the corresponding equations of an Eckart fluid. We did not need to
exploit any special additional stability conditions beyond the well
known thermodynamic inequalities and the stability conditions of
fluids. This is in strong contrast to the Israel-Stewart theory,
where one should assume additional conditions \cite{HisLin87a}.

\section{Conclusion}

In this paper we addressed causality and stability, the two most
important conceptual issues in relativistic hydrodynamics.

We have collected arguments that in dissipative first order
relativistic fluids applied to heavy ion collisions acausality
related problems may be beyond the range of validity of the theory.
Therefore first order theories, parabolic and mixed parabolic
hyperbolic equations can be useful physical models in relativistic
hydrodynamics.

Moreover, with a proper distinction of the total and internal
energies we suggested a simple modification of the Eckart theory and
we proved that its homogeneous equilibrium is stable by linear
perturbations in case of the Eckart choice of the velocity field.
With the Landau-Lifshitz form of the velocity field our theory
simplifies to the Eckart theory, therefore the corresponding
homogeneous equilibrium is stable, too \cite{KosLiu00a}. However,
the Landau-Lifshitz convention has some other undesirable
properties, that we intend to discuss in a consequent paper.

The suggested relativistic form of the internal energy depends on
the momentum density, therefore the entropy function is the function
of the momentum density as well. Moreover, in the first
approximation we have a regular second order theory with only one
additional quadratic term in the entropy four vector. However, there
was no need to introduce additional parameters, the coefficient of
the quadratic term, and therefore the relaxation time in the
generalized Fourier equation, is fixed. It was very important, that
the general entropy vector is not a simple quadratic function in the
heat flux. We have got a correction to the viscous bulk pressure,
too.

\section{Acknowledgment}

This work has been supported by the Hungarian National Science Fund
OTKA (T49466, T48489) and by a Bolyai scholarship of the Hungarian
Academy of Sciences for P.~V\'an. Enlightening discussions with
L\'aszl\'o Csernai and Hans Christian \"Ottinger are gratefully
acknowledged. {\em Mathematica} 6 was helpful in the calculations.

\bibliographystyle{unsrt}

\end{document}